\documentstyle[epsf,prd,aps,floats]{revtex}

\newcommand{\f}{\begin{equation}} 
\newcommand{\ff}{\end{equation}}
\newcommand{\be}{\begin{equation}}   
\newcommand{\ee}{\end{equation}}   
\newcommand{\bea}{\begin{eqnarray}}   
\newcommand{\eea}{\end{eqnarray}}

\begin{document}

\twocolumn[\hsize\textwidth\columnwidth\hsize\csname@twocolumnfalse\endcsname  \author{Jo\~ao Maguejo${}^\#$ and Lee Smolin${}^{* \#}$}
\address{ ${}^\#$Theoretical Physics, 
The Blackett Laboratory, Imperial 
College,  Prince Consort Road, London SW7 2BZ, UK}
\address{${}^*$Perimeter Institute for Theoretical Physics, 
Waterloo, Canada  N2J 2W9\ \ and Department of Phyiscs, 
University of Waterloo}
\title{Lorentz invariance with an invariant energy scale}
\maketitle    
\begin{abstract}
We propose a modification of special relativity in which a physical 
energy, which may be the Planck energy, joins the speed of light as an 
invariant, in spite of a complete relativity of inertial frames and 
agreement with Einstein's theory at low energies. This is 
accomplished by a non-linear modification of the action of the Lorentz 
group on momentum space, generated by adding a dilatation to each 
boost in such a way that the Planck energy remains invariant.
The associated algebra has unmodified structure constants, and we 
highlight the similarities between the group action found and 
a transformation previously proposed by Fock. 
We also discuss the resulting modifications of field theory and 
suggest a modification of the equivalence principle which determines 
how the new theory is embedded in general relativity.  
\end{abstract} 
\pacs{PACS Numbers: *** } 
 ]

\renewcommand{\thefootnote}{\arabic{footnote}} 
\setcounter{footnote}{0}

A simple paradox confronts us as we seek the quantum theory of
gravity.  The combination of gravity ($G$), the quantum ($\hbar$)
and relativity ($c$)
gives rise to the Planck length,
$
l_P = \sqrt{\hbar G / c^3 }
$
or its inverse, the Planck energy $E_P$.
These scales mark thresholds beyond which the old description of 
spacetime breaks down and qualitatively new phenomena are expected to 
appear. Thanks to the progress made by several different approaches 
to quantum gravity we have predictions for these new phenomena, which 
include discrete spatial and causal structure,  discrete spectra for 
physical observables such as area and volume\cite{discrete} 
and the appearance of 
string rather than local excitations.  

However, the new theory is expected to agree with special 
relativity when the gravitational field is weak or absent, and
in experiments probing the nature of space-time at energy scales
much smaller than $E_P$. This 
gives rise immediately to a simple question:  {\it in whose reference 
frame are $l_P$ and $E_P$ the thresholds for new phenomena?}
For suppose that there is a physical length scale which measures the
size of spatial structures in quantum spacetimes such as the discrete
area and volume predicted by loop quantum gravity. Then if this scale 
is $l_P$ in one inertial reference frame, special relativity suggests 
it may be different in another observer's frame: a straightforward
implication of the Lorentz-Fitzgerald contraction.   


There are several different possible answers to these questions. One 
is that Lorentz invariance, (both global and local) is only an
approximate symmetry, which is broken at the Planck scale.  
This has been advocated by a number of physicists, and there have
been some claims that  Lorentz symmetry breaking could be observable 
in the near future (or may even already have been observed) in  cosmic 
ray spectra\cite{cosmicray} and gamma ray bursts \cite{amel,amel1,liouv}. 
However it is troubling to contemplate giving up the principles behind 
Lorentz invariance, which are the relativity of inertial frames and 
the equivalence principle.  Does incorporating the Planck scale into 
physics mean that in the end there are preferred states of rest and 
motion?

In this letter we show that the answer is no. It is in fact possible to modify 
the action of the Lorentz group on physical measurements so that a 
given energy scale, which we will take to be the
Planck energy, is left invariant.   
That is, we can have the complete relativity of inertial 
frames and at the same 
time have all observers agree that the scale on which a transition 
from classical to quantum spacetime takes place is the Planck scale, 
which is the same in every reference frame.  
At the same time, the familiar and well 
tested actions of the boosts are maintained at large distance
or low energy  scales.
This is achieved not by a quantum deformation of the Lorentz or 
Poincare group, but by a modification of the action of the Lorentz 
group acting on momentum space. The action is defined to be 
non-linear in general, but to reduce to the usual linear action at 
energies much below the Planck scale. The non-linearities are chosen
so that the Planck energy becomes an invariant. The speed of light is 
still meaningful, and is still an invariant.


A similar proposal was made by Fock\cite{fock},
motivated by the search of the general symmetry group preserving
relativity without assuming the constancy of $c$. However in 
that case the action of the transformations are modified at large 
distances rather than large momentum.  One can understand our 
proposal as an application of the Fock-Lorentz symmetry to 
momentum space. The fact that we may preserve the invariance of the
speed of light, if we wish, is an added bonus of our approach.



Our argument is based upon four basic principles. First we assume
{\it the relativity of inertial frames}:  when 
gravitational effects can be neglected, all observers in free, 
inertial motion are equivalent. This means
there is no preferred state of 
 motion, so velocity is a purely relative quantity.
Secondly we assume {\it the equivalence principle}:  
under the effect of gravity freely falling observers are 
all equivalent to each other and are equivalent to inertial observers. 
We then introduce a new principle: 
{\it the observer independence of the Planck energy:} 
all observers agree that there is an invariant energy scale, which we 
take to be the Planck scale $E_{P}$. This will lead to novelties, but
we finally impose
{\it  the correspondence principle:} at energy scales much 
smaller than $E_{P}$ conventional special and general 
relativity are true, that is they hold to first order in
the ratio of energy scales to $E_{P}$.

The first and fourth principle tell us that there is a 
transformation group that converts measurements made by one inertial 
observer to measurements made by another.  For energy scales much 
smaller than $E_{P}$ this action should reduce to the ordinary Lorentz 
group.  Thus we expect that the Lorentz group should be replaced by 
a deformed or modified group, acting on momentum space.  
As in ordinary special relativity that group must be a six parameter 
extension of the spatial rotations group - three parameters for 
rotations and three for boosts.  
However, the only six parameter group that has these 
characteristics is the Lorentz group itself.  But we know that the 
usual linear action of the Lorentz group on momentum space does not 
fix any energy scale, as required by our third principle.  
The only possibility then is that the symmetry
group is the ordinary Lorentz group, but it acts {\it non-linearly}
on the momentum space.  That non-linear action should involve 
the 
Planck energy in some way that ensures that the Planck energy is 
preserved.
One way to do this is to combine each boost with a 
dilatation. The dilatation must be chosen so as to bring one energy 
scale 
back to the value it had before the boost transformation.
We show how to do this first for the Lorentz algebra, then for the 
Lorentz group.

Momentum space $\cal M$ is the four dimensional vector space
consisting of momentum vectors $p_a$.  The ordinary Lorentz 
generators act as
\f
L_{ab} = p_a {\partial \over \partial p^b} - 
 p_b {\partial \over \partial p^a}
\ff
where we assume a metric signature $(+,-,-,-)$
and that all generators are antihermitian (also
$a,b,c,=0,1,2,3$, $i,j,k=1,2,3$, and $c=1$).
In addition the dilatation generator $D = p_a {\partial \over \partial p_a}$
acts on momentum space as
$D \circ p_a = p_a$.
We may consider now the modified algebra, generated by the usual
rotations
$
J^i \equiv  \epsilon^{ijk} L_{ij} = \epsilon^{ijk} M_{jk}
$
and a modified generator of boosts, 
\f
K^i \equiv L_0^{\ i} + l_P p^i D \equiv M_0^{\ i}.
\label{newboosts}
\ff
We note that despite the modification, $J^i$ and $K^i$
satisfy  precisely the ordinary Lorentz algebra:
\f\label{comuts}
[J^i, K^j]= \epsilon^{ijk} K_k; \   
[K^i, K^j]= \epsilon^{ijk} J_k
\ff
(with $[J^i, J^j]= \epsilon^{ijk} J_k $ trivially preserved).
However the action
on momentum space has become non-linear due to the term in $p^i$ in
(\ref{newboosts}).  
The new action can be considered to be a 
non-standard, and non-linear embedding of the Lorentz group in the 
conformal group. 

To exponentiate the new action we note 
that 
\f
K^i = U^{-1} (p_0)  L_0^{\ i} U (p_0 )
\label{U}
\ff
where the energy dependent transformation $U(p_0)$ is given by
$ U(p_0) \equiv \exp(l_P p_0 D)$.
The non-linear representation is then generated by $U(p_0)$ and
we have
\f
U(p_0) \circ p_a = {p_a \over 1- l_P p_0 }
\ff
We note that $ U(p_0)$ is not unitary, so this is not a unitary 
equivalence. We also note that $ U(p_0)$ is singular at $p_0=l_P^{-1}$,
a property which signals the emergence of a new invariant.

The non-linear representation of the Lorentz group is then
given by
\f
W[\omega_{ab}]= U^{-1}(p_0)  e^{ \omega^{ab}L_{ab}}U(p_0) = 
 e^{\omega^{ab}M(p_0)_{ab}}
 \label{nonlinear}
\ff
In evaluating this expression, note that $D$ acts on everything to the 
right, and $p_0$ always means the time component of the vector 
immediately to the
right.    Using these rules, one finds that 
the boosts in the $z$ direction are now given by:
\bea \label{fltransp}
p_0'&=&{\gamma\left(p_0- vp_z \right) 
\over 1+l_P (\gamma -1) p_0  -l_P\gamma v p_z }\\  
p_z'&=&{\gamma\left(p_z- v p_0\right) 
\over 1+l_P (\gamma -1) p_0  -l_P\gamma v p_z }\\  
p_x'&=&{p_x
\over 1+l_P (\gamma -1) p_0  -l_P\gamma v p_z }\\  
p_y'&=&{p_y
\over 1+l_P (\gamma -1) p_0  -l_P\gamma v p_z }
\eea 
which reduces to the usual transformations for small $|p_\mu|$.

This transformation is identical with a transformation introduced
by Fock\cite{fock}  but applied to momentum space.
Fock's transformation is obtained from the one above replacing $p$
with $x$, and therefore its generators also satisfy the 
standard commutators (\ref{comuts}).
However non-linearity means that the group action in 
spatial and momentum space are radically different. Indeed Fock's 
transformation (defined in $x$ space) reduces to Lorentz at 
{\it small} distances (so that it defines a  {\it large} invariant Planck 
length). On the contrary, our transformation (defined in $p$ space) 
becomes Lorentz for {\it small} energies and momenta (and defines a 
{\it large} invariant Planck energy, as we shall see) - 
the property we are looking for. Also Fock's transformation contains a
varying speed of light\cite{fock,man},  whereas, as we shall see, our 
proposal does not.

It is not hard to see that the Planck energy is preserved by the 
modified action of the Lorentz group. For example, boosts in the $z$ 
direction with
velocity $v$ take
$
(E_p, 0, 0 , 0) \rightarrow (E_p, -v E_p , 0 , 0)
$.
From the group property we can also deduce (and then check)
the following. Suppose we observe a particle in our frame
with energy momentum, $(E_p, P , 0 , 0)$ with $P/E_p < 1$.
Then a boost in the $-\hat{z}$ direction with $v= P/E_p$ will
bring us to the Planck mass particle's rest frame with
$
(E_p, P, 0 , 0) \rightarrow (E_p, 0 , 0 , 0) .  
$
Furthermore the $4$-momenta of photons with Planck energy 
$E_p$ traveling in the $z$ direction are preserved under
boosts in the $z$ direction, because
$
(E_p, E_p , 0 , 0) \rightarrow (E_p, E_p , 0 , 0)
$.

Clearly these transformations do not preserve
the usual quadratic invariant on momentum space. But there is a 
modified invariant, obtained from $U( \eta^{ab}p_a p_b )$,
which is:
\f\label{invariant}
||p||^2 \equiv {\eta^{ab}p_a p_b \over (1-l_P p_0 )^2}
\label{inv}
\ff
This invariant is infinite for the new invariant energy scale
of the theory $E=l_P^{-1}$, and it's not quadratic for energies close or above 
$E=l_P^{-1}$. This signals the expected collapse in this regime of 
the concept of metric (i.e. a quadratic invariant).

It is also evident from (\ref{inv}) that the symmetry of positive
and negative values of the energy is broken.  The formalism
may be defined with $l_P$ equal to minus the Planck length, in which case the
invariant diverges for energy $E=-E_p$.  The two theories with the
two signs of $l_P$ are physically distinct; and we know of no
theoretical consideration which fixes the sign of $l_P$. 
Even though in what follows we shall assume $E_P>0$,
we will also briefly consider how conclusions change if $E_P<0$, so that both 
the sign and magnitude of $l_P$ may be determined experimentally,
from effects that we shall now discuss.

We  start by considering massive particles.
These have a positive invariant
$
||p||^2 >0
$
which may be identified with the square of the mass
$||p||^2 =m_0^2c^4$. Considering the rest frame we therefore
obtain a modified relation between energy and mass:
\be
E_0={m_0c^2\over 1+{m_0c^2\over E_p}}
\ee
In a general frame we find that $m$ transforms in the usual way 
$m=\gamma m_0$, however:
\bea
E&=&{m\over 1+{m\over E_p}}\\
p&=&{mv\over 1+{m \over E_p}}
\eea
It is at once obvious that the energy of a particle can never 
equal or exceed $E_p$, even though its mass may be as large as wanted.
Asymptotically a particle may have $E=E_p$ if it has infinite
rest mass. Its energy and momentum are then frame independent,
in agreement with the postulates of the theory. 
Notice that if $E_P>0$ the energy of a particle is smaller than 
the usual $E=mc^2$; however if $E_P<0$ its energy is larger than
$mc^2$  and in fact diverges for Planck mass
particles.

All these remarks apply to fundamental particles, not macroscopic
sets of them. The latter may have masses larger than $E_P$, but if
they are made of particles with $E\ll E_P$ they do not feel 
the transformations (\ref{fltransp}) because these, being non-linear,
are not additive.

The modified invariant for photons still has the property:
$
||p||^2 =0
$
and so $E=p_0=|p_i|$. Consider a photon moving in the $z$ direction,
so that $E=|p_i|=p_z$, and consider a boost in the $z$ direction
as above. We thus obtain the Doppler shift formula  
\be
E'={E\gamma (1-v)\over 1+(\gamma(1-v)-1)l_PE}
\ee
This can be rewritten as 
\be\label{ens}
{1\over E'}-{1\over E_P}={1\over \gamma (1-v)}{\left(
{1\over E}-{1\over E_P}\right)}
\ee
showing how $E=E_p=1/l_P$ is invariant - so the Planck energy
and momentum for photons is frame independent. 
Furthermore super and sub Planckian energies never get mixed via 
Doppler
shift, as $ \gamma (1-v)>0$ and the sign of both sides of 
eqn.~(\ref{ens}) 
must  be the same. It is impossible to blueshift a sub-Planckian 
photon up to $E_p$, or redshift  a super-Planckian photon down
to $E_p$. Closer inspection reveals an abnormality:
super-Planckian photons redshift if the source moves towards
the observer, blueshift otherwise. It is impossible to redshift
them below $E_p$ whatever the speed of a source towards us.
If the source moves away from us there
is a recession speed for which $E'=\infty$ and beyond which 
$E'<0$. (These remarks apply to $E_P>0$ only).

Using the equivalence principle we can now derive a 
formula for the  first order gravitational redshift.
In the non-relativistic regime the Doppler shift is
${\Delta E\over E}=v{\left(  1-l_PE\right)}$
showing a decrease in the Doppler shift as the photon
approaches the Planck energy. Using the equivalence 
principle 
this translates into a similar modification
for the gravitational shift
\be
{\Delta E\over E}=\Delta \phi{\left(  1-l_PE\right)}
\ee
The Pound Rebbka experiment is of course not sensitive enough
for detecting this new effect, but ultra high energy 
cosmic rays (UHECR) might not be.

In future work we shall examine the effects of our proposal
for fields of all spins, including gravity. Here we merely outline
how our approach leads to modifications in field theory, considering
the case of a scalar field.
Up till now we considered the 
modification of the Lorentz transformations on momentum space. When 
applied to field theory, the derivatives of a field should transform 
as momentum, as they correspond to physical frequencies and 
wavelengths.
Thus, under a change of inertial observers we have
$
(\partial_a \phi ) \rightarrow (\partial_a \phi )^\prime =
W(\partial_0 \phi ) _a^{\ b}  (\partial_b \phi ) 
$
and we see that the transformation is non-linear, with
$l_P \partial_0 \phi$ playing the role of $p_0$. 

The action for a scalar field must be Lorentz invariant, but in the 
present context this means that it should be invariant under the 
modified Lorentz transformations.  The invariant action is then
\f
S^\phi = \int d^4x {1\over 2} {\eta^{ab} (\partial_a \phi ) 
(\partial_b \phi ) 
\over [1 - l_P^2 (\partial_0 \phi ) ]^2 } + {m^2 \over 2} \phi^2
\ff
Because there 
is now no quadratic invariant, there is no linear field equation. 
Instead we may derive a complex non-linear field equation, as we shall
do in a future publication. The surpising thing is that a single plane wave
$
\phi_k (x) = A \exp(-\imath k_a x^a)
$
is still a solution with $\eta^{ab}k_ak_b =0$, if $m=0$. 
However the superposition
principle no longer holds.  It is also the case that 
for massive fields
there no longer is an exact plane wave solution.

To extend the new theory to a modification of general relativity we 
must find the appropriate way to express the equivalence principle.
Given that we have modified the action of the Lorentz transformations
in special relativity in the momentum space we proceed by remarking
the following:  1)  Matter is most 
generally represented in general relativity in terms of fields;
2) Momenta of fields is associated with spatial derivatives; 
3) In a field theory, the mathematical tangent space 
corresponds physically most closely to the derivatives of fields;
4) By the equivalence principle, the Lorentz group acts on components of
fields referred to orthonormal frames.
This leads us to a
{\it modified equivalence principle:} the non-linear realization of the
Lorentz group discussed above acts on derivatives of fields referred
to orthonormal frame components. That is, in the presence of gravity
the transformations proposed are
defined for  quantities of the form 
$
(\partial_a \phi)|_x \equiv e_a^\mu (p) (\partial_\mu \phi)_x
$
(where greek letters refer to ordinary manifold 
coordinates and latin letters to components 
in an orthonormal frame).
These transform according to the non-linear realization, i.e.
measurements made by two orthonormal frames,
$e_a^\mu$ and $e_a^{\prime \mu}$, of derivatives of a scalar  
field are related by
$
(\partial^\prime_a \phi) = W(\partial_0 \phi)_a^b (\partial_b \phi )
$
where $W(\partial_0 \phi)_a^b$ is defined
by (\ref{nonlinear}) and depends on $\partial_0 
\phi|_p$ at event $p$ in the same way that the momentum space
realization (\ref{nonlinear}) depended on the energy $p_0$.

We see that the orthonormal frame components themselves do not
have well defined transformation rules under these modified
transformations.  We consider these abstract mathematical
quantities, while our transformation rule only applies to physical
measurements of momenta. 
Similarly, there is no new transformation rule for the manifold
derivatives $\partial_\mu \phi$ as these also do not relate
to measurements made by freely falling observers. The latter
are described by $\partial_a \phi$ and so it is only to these,
and not to their separate mathematical parts, that the
new transformation rules apply.  
In future work we show that this prescription fully defines how to
incorporate our transformation into field theory for all spins. 

In summary what we have proposed here is a modification of the two basic 
principles of physics: the relativity of inertial frames and the 
equivalence principle. The modifications proposed are the simplest 
ones we are aware of consistent with the demand that the Planck 
energy be an invariant, while special relativity as formulated by
Einstein hold at much lower energies. We explored implications for
well known special relativistic effects, and found modified formulae
for them.

We close this letter with  a number of questions, to be examined in 
future work.
Foremost: does the modified equivalence 
principle we have just stated lead uniquely to a consistent modification 
of general relativity? The fact that the algebra of the symmetry group 
remains the same suggests that perhaps the standard spin connection 
formulation of relativity is still valid. At high energies there is no 
longer a metric, as the invariant (\ref{invariant}) 
is no longer quadratic. However the connection, taking values in the 
algebra, is still unmodified, and one may define curvature and the usual 
tools of Riemanian geometry without any trouble. We hope to return to
this issue and study implications for cosmology and 
black hole physics \cite{bh}.
  
One may further ask 
how the modified action of the Lorentz group 
is to be extended to spinor fields? A related issue is whether 
supersymmetry can be modified to be consistent with the
modified action. Does this lead to mass 
differences between supersymmetric partners?  Can string theory 
also be modified to be consistent with the principles described 
here? Could the principles proposed here be derived from the 
large distance limit of causal spin foam models, which incorporate 
discrete spatial and causal structure at the Planck scale?

We finally note that we can find many non-linear
realizations of the action of the Lorentz group, by making
other choices for $U(p_0)$ in eq. (\ref{U}). These lead to
other forms for the  modified invariants and hence to different dispersion 
relations for massive and massless particles. It is interesting to
ask to what extent these can be distinguished experimentally by data from 
gamma ray bursts and UHECRs? More general choices of $U(p_0)$ in eq. (\ref{U})
in general lead to invariants which contain an energy
dependent speed of light. Could these theories be used to 
implement the varying speed of light cosmology\cite{mof1,am}?

{\bf Acknowledgements } We are grateful for conversations 
with  S. Alexander,  G. Amelino-Camelia,
J. Kilowski-Gilkman, F. Markopoulou, and A. Tseytlin, which have helped us 
to understand better the idea proposed here.  
LS was supported by the NSF through grant
PHY95-14240 and a gift from the Jesse 
Phillips Foundation.

\end{document}